\documentclass[aps,pre,reprint,groupedaddress,amssymb]{revtex4-2}

\makeatletter
\def\set@curr@file#1{%
  \begingroup
    \escapechar\m@ne
    \xdef\@curr@file{\expandafter\string\csname #1\endcsname}%
  \endgroup
}
\def\quote@name#1{"\quote@@name#1\@gobble""}
\def\quote@@name#1"{#1\quote@@name}
\def\unquote@name#1{\quote@@name#1\@gobble"}
\makeatother
\usepackage{graphicx}
\usepackage{dcolumn}
\usepackage{bm}
\usepackage{amsmath}
\usepackage{xcolor}
\usepackage{caption}
\usepackage{subcaption}
\usepackage{epstopdf}
\usepackage{hyperref}
\usepackage{ulem}
\usepackage[utf8]{inputenc}
\usepackage[margin=10pt,font=small,labelfont=bf,justification=centerlast,format=plain]{caption}
\usepackage{lipsum}

\begin{document}


\title{Competitive balance theory: Modeling conflict of interest in a heterogeneous network}
\author{F. Oloomi\textsuperscript{1}}
\author{R. Masoumi\textsuperscript{1}}
\author{K. Karimipour\textsuperscript{1,2}}
\author{A. Hosseiny\textsuperscript{1}}
\email{al\_hosseiny@sub.ac.ir}
\author{G. R. Jafari\textsuperscript{1,3}}
\email{g\_jafari@sbu.ac.ir}	

\affiliation{\textsuperscript{1} Department of Physics, Shahid Beheshti University, G.C., Evin, Tehran 19839, Iran}
\affiliation{\textsuperscript{2}Department of Sociology, McMaster University, 1280 Main St W, Hamilton, Ontario, L8S 4L8, Canada}
\affiliation{\textsuperscript{3}Department of Network and Data Science, Central European University, H-1051 Budapest, Hungary}

\date{\today}

\begin{abstract}
The dynamics of networks  on Heider’s balance theory moves toward reducing the tension by constantly reevaluating the interactions to achieve a state of balance. Conflict of interest, however, is inherent in most complex systems; frequently, there are multiple ideals or states of balance, and moving towards one could work against another. In this paper, by introducing the competitive balance theory, we study the evolution of balance in the presence of conflicts of interest. In our model, the assumption is that different states of balance compete in the evolution process to dominate the system. We ask, whether, through these interactions, different states of balance compete to prevail their own ideals or a set of co-existing ideals in a balanced condition is a possible outcome. The results show that although there is a symmetry in the type of balance, the system either evolves towards a symmetry breaking where one of the states of balance dominates the system, or, less frequently, the competing states of balance coexist in a jammed state. 
\end{abstract}


\maketitle

\section{Introduction}

Conflict of interest is frequently observed in social networks, with conflicting sources of authority pushing the network towards higher representation. Assuming that a social system changes and evolves towards balance, these different sources of authority compete for a higher share. As the system changes and evolves towards balance, these different ideals compete to dominate the network.  In this paper, we aim to extend the structural balance model, as theorized in Heider’s pioneering work \cite{heider1946}, to take this competition of ideals into account. 
A half century of research has shown that balanced networks have been very well-theorized \cite{park2005, meghdad2017, kirkley2018, rabbani2019, parravano2016} and are frequently observed empirically \cite{hart1974, hummon2003, Lerner2016, leskovec2010, Doreian2015}. The model was first introduced by Heider and then modeled in signed graphs by Cartwright and Harary \cite{Cartwright1956}, and developed to explain the structure of conflicts and balance in a signed network; a network in which the connections are characterized as friendship and hostility, denoted by $\pm 1$. A triad is balanced when all three agents are friends, or when two friends have a common enemy; the cycle, otherwise, is unbalanced. Overtime, the system either reaches a global minimum (either all links become positive or two hostile clusters emerge) or gets trapped in a local minimum (jammed state).

This original form of structural balance theory has been extended in several directions \cite{Belaza2017, Du2018, Belaza2019,Kargaran2020}. For example, Antal, Krapivsky, and Redner \cite{Antal2005}  go beyond a static description of balanced relations, and investigate different dynamical rules \cite{Kulakowski2005, Marvel2011, Abell2009, Leila} for achieving balance. Also, Marvel, Strogatz, and Kleinberg \cite{Marvel2009}, on the basis of social psychology theories, suggest that certain triads are more stable than others, and taking this statement as the driving force for change, they extract the energy landscape of the networks.

Heider’s balance theory, in its original form, models the evolution of a network towards balance, using just one type of link. Conflict of interest is inherent in most complex systems; frequently, there are multiple states of balance, and moving towards one could work against another. In this paper, we extend the model to propose the competitive balance theory: evolving towards balance in the presence of conflict of interest, represented by introducing several different types \cite{Kulakowski2017, forogh2017, Sheykhali2019} of $\pm$ of competing links. In our model, different types of links represent different conceptualizations of friendship and hostility.

We can employ the social scientific concept of "conflicting discourses" to clarify the idea of conflict of interest and different conceptualizations of friendship and hostility (i.e. different types of $\pm$ links). Discourse, as a social scientific term and as elaborated by Foucault \cite{Foucault1980}, in this context, refers to socially shared habits of thoughts, perceptions, and behaviors. The dominant discourse is attached to power, is continuously reaffirmed as the normative discourse, and often involves ideas about a particular group of people [e.g.  Jews, women or the LGBTQ community (Lesbian, Gay, Bisexual, Transgender, Queer or Questioning)]. As the central point of commonality, the dominant discourse is the accepted way of perceiving social reality, or behaving in social space. Alternative discourses, are usually those of the non-power holding others and are then shaped in response to the dominant discourse, frequently at the margins of systems, especially at the beginning of their emergence \cite{Hall2001, Spargo1999, Samin}.  Discourses are not attributable to individuals; they are a property of communication. That is why in our model discourses are represented as the attributes of edges (different types of edges) as opposed to nodal attributes. The discourse in which the interaction between two individuals is taking place can change, and this is denoted through changing the type of link.

This extended theory of structural balance can be used to model the dynamic behavior of the diverse, interacting, and possibly competing, discourses that coexist in a system. The competitive balance theory acknowledges that the meanings of hostility and friendship are localized in a social space, and different discourses, with potential or perceived conflicts of interest, compete to reach balance or dominance.
Conflict of interest frequently arises between the dominant and alternative discourses. In the global environment, with several discourses operating at the same time, the meanings of friendship and hostility (links) are localized. Friendship and hostility are defined based on the discourse in which the interactions are taking place, with each of the active discourses having its own conceptualization of links. Discourses are also in dialogue all the time; they are rarely, if ever, independent from one another or entirely cut off from the global environment. Although each discourse has its own internal conflicts, conflicts of interest among discourses to dominate the system create a conflictual and contentious social space. In short, in our model, there are various and potentially competing forms of links and different desired balanced states. 

Considering the inter- and intra discourse interactions, the question is, whether, through these exchanges, the competition evolving towards homogenization of a type of link (i.e., one discourse privileging its own set of meanings) or, alternatively, a set of coexisting discourses in a balanced condition (i.e., coexisting states of balance) is a possible outcome.

\section{\label{sec2}Model }

As explained in the introduction, in the real world, the meanings of hostility
and friendship are localized in social space; in other words, people have different sets of reasons to be friends or resent one another. Taking this heterogeneity into account, the state of balance or unbalance in a triad is influenced by these different meanings of friendship and hostility.

Since in the original conceptualization of a balanced triad the assumption is that there is a single meaning (or cause) of friendship or hostility, all triads are
"pure". In other words, going back to the concept of discourse, all dyadic relationships in a triad function are in the same discourse, or are derived from the
same order of meanings. Extending the model to include diverse systems of meanings, in addition to the pure triads (getting their meanings from a
single discourse), there will be "mixed" ones with edges coming from different
realms of meanings. 

\begin{figure}[!h]
     \centering
      \includegraphics[width=1\linewidth]{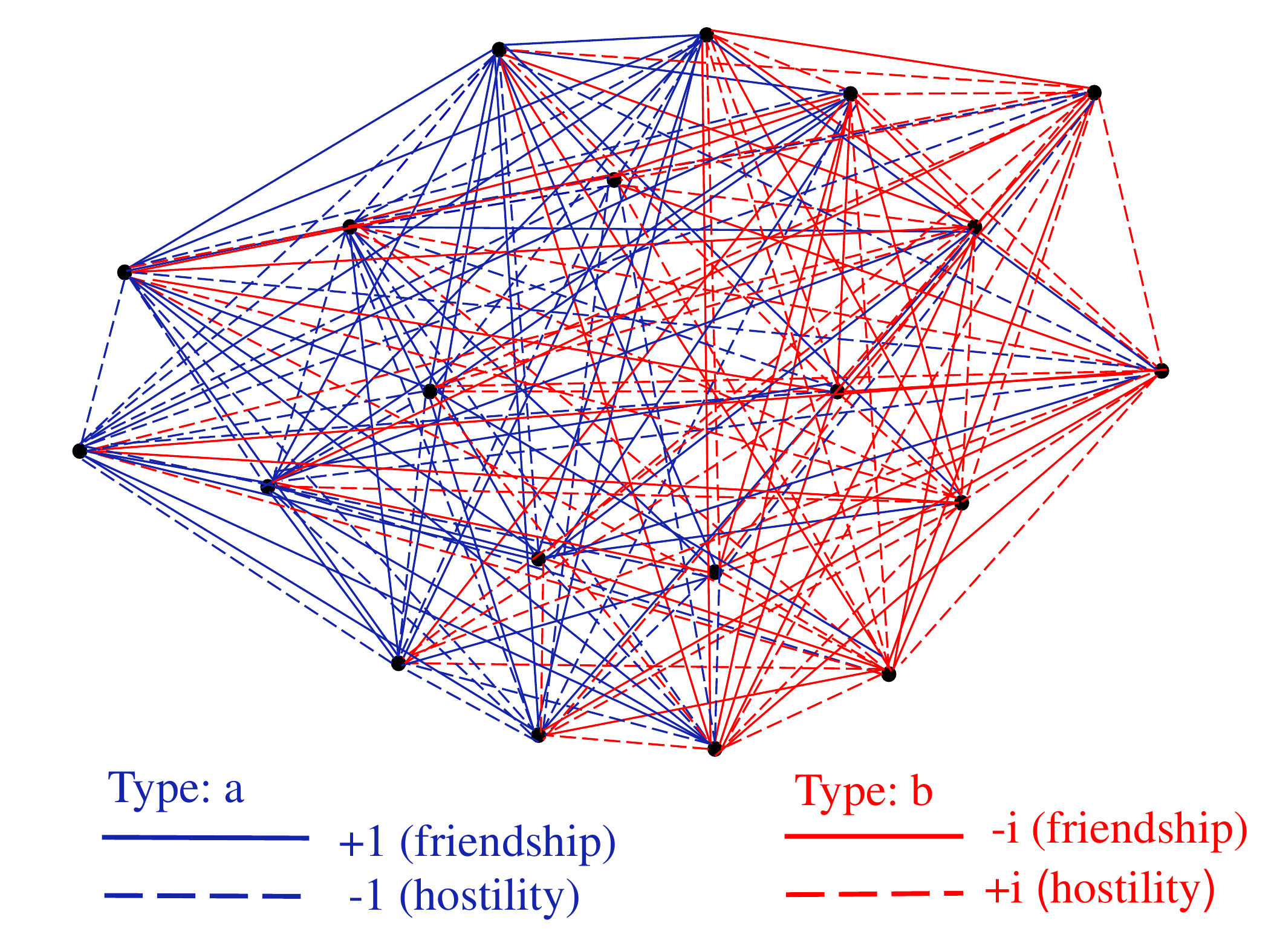}
	\caption{ A network with two different conceptualizations of friendship and hostility. The solid and dashed lines show friendship and hostility, respectively, and the colors refer to the types of links (real or imaginary).}
	\label{Fig1}
\end{figure}

In this model, for simplicity, we use two different conceptualizations of
friendship and hostility, i.e., four types of links altogether $\pm a$ and $\pm b$.
Fig. \ref{Fig1} is a proper representation of such a network. The $\pm a$ edges aim to
evolve towards the first desired balanced state, and the $\pm b$’s aim to
evolve towards the second desired balanced state, with a conflict of interest
occurring among $\pm a$’s and $\pm b$’s.

We, then, pursue to model the presence of two different types of
friendship and hostility, satisfying the following conditions:
\begin{enumerate} 
	\item {All the triads (pure and mixed) are homogeneously treated and their
	energies are equally weighted in our Hamiltonian function. The triads are
	homogeneous if, in moving towards lesser tension, the system does not
	prefer one balanced triad over another (i.e. the pure ones over the mixed
	ones, or the other way). In other words, the system moves towards the
	presence of more balanced triads, irrespective of the dyadic discursive
	relationships involved in each.}
	\item {The definitions of balanced and unbalanced pure triads follow the original
	conceptualization by Heider \cite{heider1946}.}
	\item  {Mixed triads, the ones with dyadic relationships stemming from different
	discursive spaces, follow a symmetrical behavior according to their energies.}
	\item  {The edges and triads that are derived from different discursive spaces are always identifiable, and to this end we need to use a label to distinguish them.}
\end{enumerate}

To achieve these, we use the properties of complex numbers. Based on this choice, the edges of type $a$ (belonging to the first order of meanings) are denoted with real numbers $\pm 1$ and the edges of type $b$ (belonging to the second order of meanings), are denoted by the imaginary numbers $\pm i$, as $+1$ and $-i$ indicate friendship and $-1$, $+i$ indicate hostility. We consider $- i$  and $+i$ as friendship and hostility, respectively, for type $b$, because according to the structural balance theory, multiplication of three friendly (hostile) edges in a pure triad is positive (negative), i.e., $(-i)^3 = +i$ or $(+i)^3=-i$.

\begin{figure}[!ht]
      \includegraphics[width=1\linewidth]{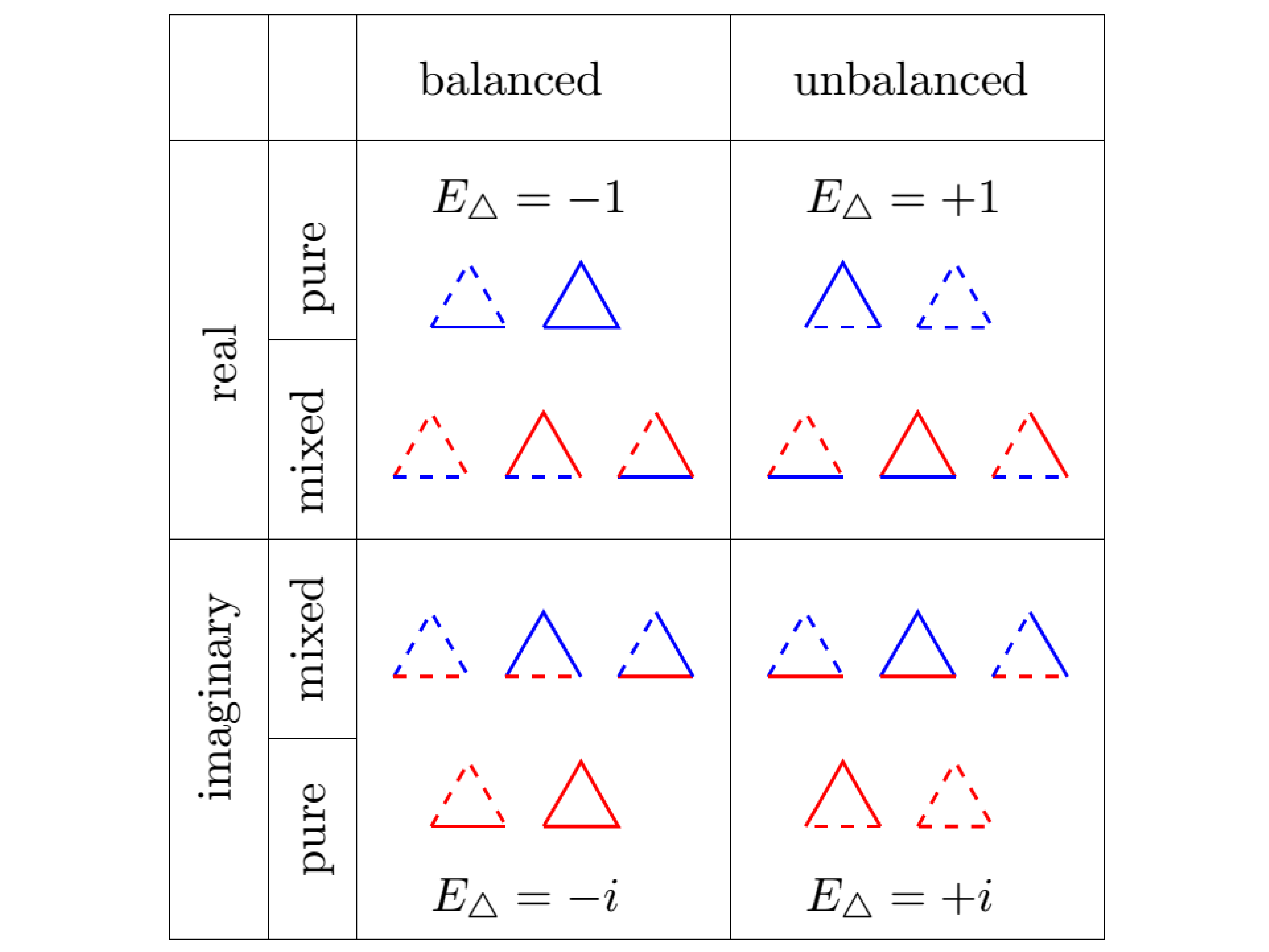}\hfill 
	\caption{All 20 possible kinds of triads are categorized based on the types and signs of their links.
		The solid and dashed lines show friendship and hostility, respectively, and the colors refer to the types of links (real or imaginary).
	}
	\label{Fig2}
\end{figure}

Fortunately, by using complex numbers, we can distinguish between the two different types of friendship and hostility without disturbing the homogeneity of triads; i.e.,
using this method the energies of pure and mixed triads are treated equally in the Hamiltonian function.
Translated into the mathematical language of modeling:

\begin{eqnarray}
\label{eq:1}
\centering
\forall \;  i , j  \in \{1, 2, ..., N\} : \; S_{ij}\in \{+1,-1, +i,-i\},
\end{eqnarray}
thus the possible energies for triad $ijk$ are as below:
\begin{eqnarray}
\label{eq:2}
\centering
E_{\triangle_{ijk}}=-S_{ij} S_{jk} S_{ki} \; \; \Rightarrow \;	E_{\triangle_{ijk}}\in \{+1, -1, +i, -i\}.
\end{eqnarray}
Indeed, based on these two types and two signs of edges, 20 different kinds of triads can be constructed as shown in Fig. \ref{Fig2}. These triads are divided into four categories according to their energy. A triad is balanced if its energy is negative and unbalanced otherwise. On the other hand, a triad is classified in the real or imaginary group if its energy is a real or imaginary number. 
\begin{eqnarray}
\label{eq:3}
\mathbf{\triangle_{ijk}=\begin{cases}
\triangle_{re}, & \text{if \; $E_{\triangle_{ijk}}=\begin{cases}
	-1, & \text{ balanced}\\
	+1, & \text{ unbalanced}
	\end{cases}$}\\ \\
\triangle_{im}, & \text{if \; $E_{\triangle_{ijk}}=\begin{cases}
	-i, & \text{ balanced}\\
+i, & \text{ unbalanced}
	\end{cases}$}
\end{cases}}
\end{eqnarray}

As mentioned before, in each group, we name a triad pure if its sides are all of the same type, either real or imaginary, and mixed if its sides are a combination of both types.

In the initial condition of simulation, we assume that all four types of edges are randomly distributed with an equal probability in a fully connected network. Since a random configuration is obviously not balanced, the network begins to evolve. As Antal, Krapivsky and Redner have mentioned the natural tendency of members in a social network is to reduce the unbalanced triads \cite{Antal2005} which is equal to tension reduction. So, in order to investigate the final states of the system, we define a Hamiltonian function that satisfies the tension reduction, as follows:

\begin{eqnarray}
\label{eq:4}
\centering
E(t)=E_{re}(t)+E_{im}(t),
\end{eqnarray}
where $E_{re}$ and $E_{im}$, respectively, are the average energy of all real and imaginary triads: 
\begin{eqnarray}
\label{eq:5}
\centering
E_{re}(t)= \frac{1}{N_{\triangle}}\sum_{\triangle \in \triangle_{re}}{E_{\triangle}(t)}\; \; \; \Rightarrow \; -1\leq E_{re}(t) \leq 1 ,
\end{eqnarray}

\begin{eqnarray}
\label{eq:6}
\centering
E_{im}(t)= \frac{1}{iN_{\triangle}}\sum_{\triangle \in \triangle_{im}}{E_{\triangle}(t)}\; \; \; \Rightarrow \; 	-1\leq E_{im}(t) \leq 1.
\end{eqnarray}
Although the share of $E_{re}$ and $E_{im}$  in the total energy varies in time, the total energy of the network is always  limited to a square area:
\begin{eqnarray}
\label{eq:7}
\centering
 \mid E_{re} \mid+\mid E_{im} \mid\leq 1 \; \; \; \Rightarrow \; -1\leq E(t) \leq 1.
\end{eqnarray}
This form of Hamiltonian simply applies constrained triad dynamics, in which updates that increase the number of unbalanced triads are not acceptable\cite{Antal2005}. In other words, the priority is to maximize the balanced triads, irrespective of their types. In each update step, we choose an edge at random to flip to one of the three other relations with equal probability. The flip will be accepted if the total number of balanced \cite{terzi2011} triads does not decrease.

In this way, the system dynamics moving toward structural balance, or lesser tension, can be studied when there are two interacting discourses at play. In the next section, we ask how a system evolves in the presence of two different, but structurally equal, types of friendship and hostility, when the model does not prefer one type over another; what is the share of each, $E_{re}$ and $E_{im}$, in the total energy during the evolution; and whether, as a result of the competition, one of the orders of meanings dominates the system or they both coexist in a balanced state.

\section{\label{sec4} Results }
\subsection{Energy in final states}

\begin{figure}[!h]
      \includegraphics[width=1\linewidth]{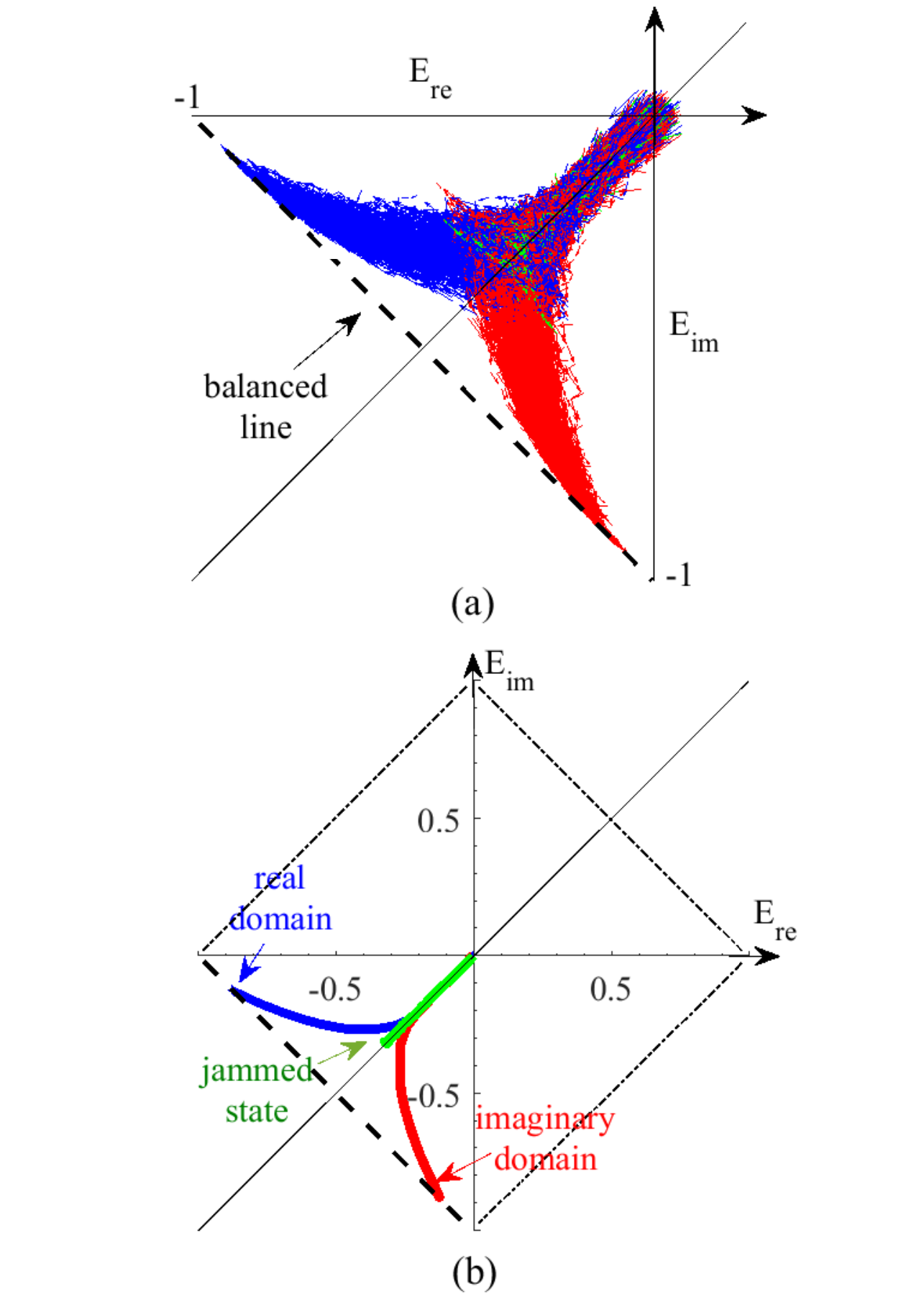}
	\caption{ The share of $E_{re}$ and $E_{im}$ in total energy during
the evolution of the network.  $E_{re}$ and $E_{im}$, respectively, show the energy of real and imaginary triads.
		 (a) The energy path of 10,000 realizations of a fully connected network with $N=20$ nodes and completely random initial condition. (b) The three main branches resulting from averaging over all paths in panel (a). }
	\label{Fig3}
\end{figure}

Fig. \ref{Fig3}a represents the share of energy of the real and imaginary triads in total energy overtime for 10,000 realizations. Note that the illustrated results are of a typical network of size $N=20$; at the end of this section we investigate the effect of size. 
At the initial state, there is no preference for different types of edges, so the densities of the four groups of triads (shown in Fig. \ref{Fig2}) are equal; hence, $E_{re}$ and $E_{im}$ have roughly the same share. The fact that at the beginning the numbers of balanced and unbalanced triads are equal implies that the starting point of the dynamics of each sample is around the origin of the coordinate. 
As time passes, the system evolves on the basis of increasing the number of balanced triads and at a specific point there appears to be a symmetry breaking. From this point forward, as shown in Fig. \ref{Fig3}a, the energy of the network falls into either the blue branch where the share of real energy is dominant or the red branch where the share of imaginary energy is dominant. 
Of course, in rare cases, the system stops in the middle of the path (green branch). 

If we let the system evolve long enough, it is finally trapped in a minimum which may be local or global. Global minima are absorbing states of the system in which $E=E_{re}+E_{im}=-1$.
 We say the system is in the real (imaginary) domain, if these two conditions are fulfilled simultaneously: $(1) E=E_{re}+E_{im}=-1$ and $(2) E_{re}>E_{im} (E_{re}<E_{im})$. However, not all the points in the $E_{re}+E_{im}=-1$ line are observed as final states and the results show that when the system reaches a balanced state in the real (imaginary) domain, the share of $E_{re} (E_{im})$ in the total energy is significantly higher than the share of $E_{im} (E_{re})$.
 It is worth noting that the symmetry is spontaneously broken, in the sense that without the intervention of an external force, in global minima the major share of energy belongs to either the real or the imaginary part.

 Also, less frequently, the system is trapped in local minima, called jammed states \cite{Antal2005}, and, as illustrated, unlike global minima, in jammed states $E_{re}$ and $E_{im}$ have close shares. If the system does not get stuck in jammed states, the energy path is finally absorbed into one of the real or imaginary domains. We averaged over the energy paths of different realizations that led to either local minima (i.e.jammed states), or global minima (i.e. real and imaginary domains, separately), as shown in Fig. \ref{Fig3}b. It is clearly evident that there is a symmetry on the path to both real and imaginary domains.

But how does the system come to prefer one domain over the other?
Although the system evolves by gradually increasing the number of balanced triads, there are different types of balanced triads that cannot appear at the same time. On one hand, as each edge is shared between the neighboring triads, not all types of balanced triads can be neighbors. Take the balanced triads $(1,1,1)$ and $(-i,-i,-i)$, for example; since they do not have an edge in common, they cannot be neighbors. 
On the other hand, even if those have a common edge, the network cannot tolerate different types of balanced states at the same time: for example, take a fully connected network of size $N=4$, containing four triads. Assume that the triad $ABD$ and $BDC$ are both balanced, respectively, with edges $(1,1,1)$ and $(i, -i, 1)$. Here, we are trying to define $AC$ in a way that the two other triads, $ABC$ and $ADC$, are also balanced. But, as shown below, with this configuration, none of the possible types of edge for $AC$ can make all four triads balanced:
\begin{enumerate} 
\item${\text{if} \; AC=+1  \Rightarrow \triangle_{ABC} \;\text{is unbalanced,} \;\triangle_{ADC} \text{is balanced}}$
       \item${\text{if} \; AC=-1  \Rightarrow \triangle_{ABC}\;\text{is balanced,} \; \triangle_{ADC}\;\text{is unbalanced}}$
       \item${\text{if} \; AC=+i  \Rightarrow \triangle_{ABC}\;\text{is balanced,}  \;\triangle_{ADC}\;\text{is unbalanced}}$
       \item${\text{if} \; AC=-i \Rightarrow \triangle_{ABC}\;\text{is unbalanced,}  \;\triangle_{ADC}\;\text{is balanced.}}$
\end{enumerate}
 As the example illustrated, not all different types of real and imaginary balanced triads can coexist in a system that has reached its final balanced state. So, there is a competition in the system; the emergence of one domain of balance results in the elimination of another.

\begin{figure}[!h]
     \includegraphics[width=1\linewidth]{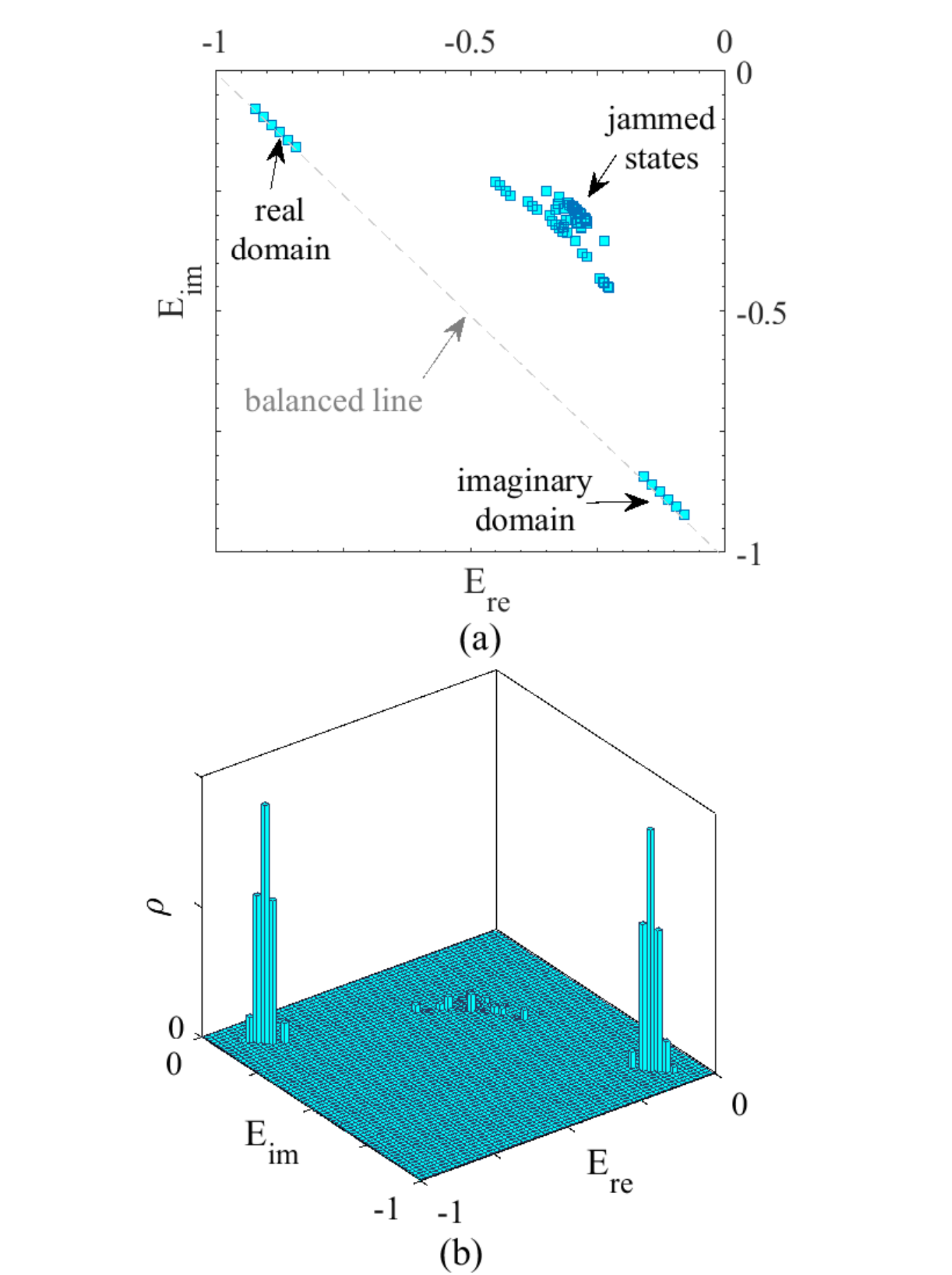}
	\caption{ Final states in different domains.
		(a) The fixed points of each domain. (b) The probability of occurrence of each fixed point.  $E_{re}$ and $E_{im}$, respectively, show the share of real and imaginary triads in total energy.}
	\label{Fig4}
\end{figure}

Fig. \ref{Fig4}a illustrates the final energy of 10,000 realizations with the same initial condition after $N^4$ Monte-Carlo steps. 
As shown, all paths are absorbed into a limited number of discrete fixed points, categorized into three sets:  jammed states, and real and imaginary domains. 
This shows that from all possibilities of balanced final states in the $E_{re}+E_{im}=-1$ line, the simulation results contain only a limited interval of discrete values. Fig. \ref{Fig4}b gives an insight about the probability of occurrence of these discrete fixed values. It shows that some values are much more likely than the others.
Also, as shown, reaching a real or an imaginary domain is significantly more probable than getting stuck in the jammed states.

\begin{figure}[!h]
      \includegraphics[width=1\linewidth]{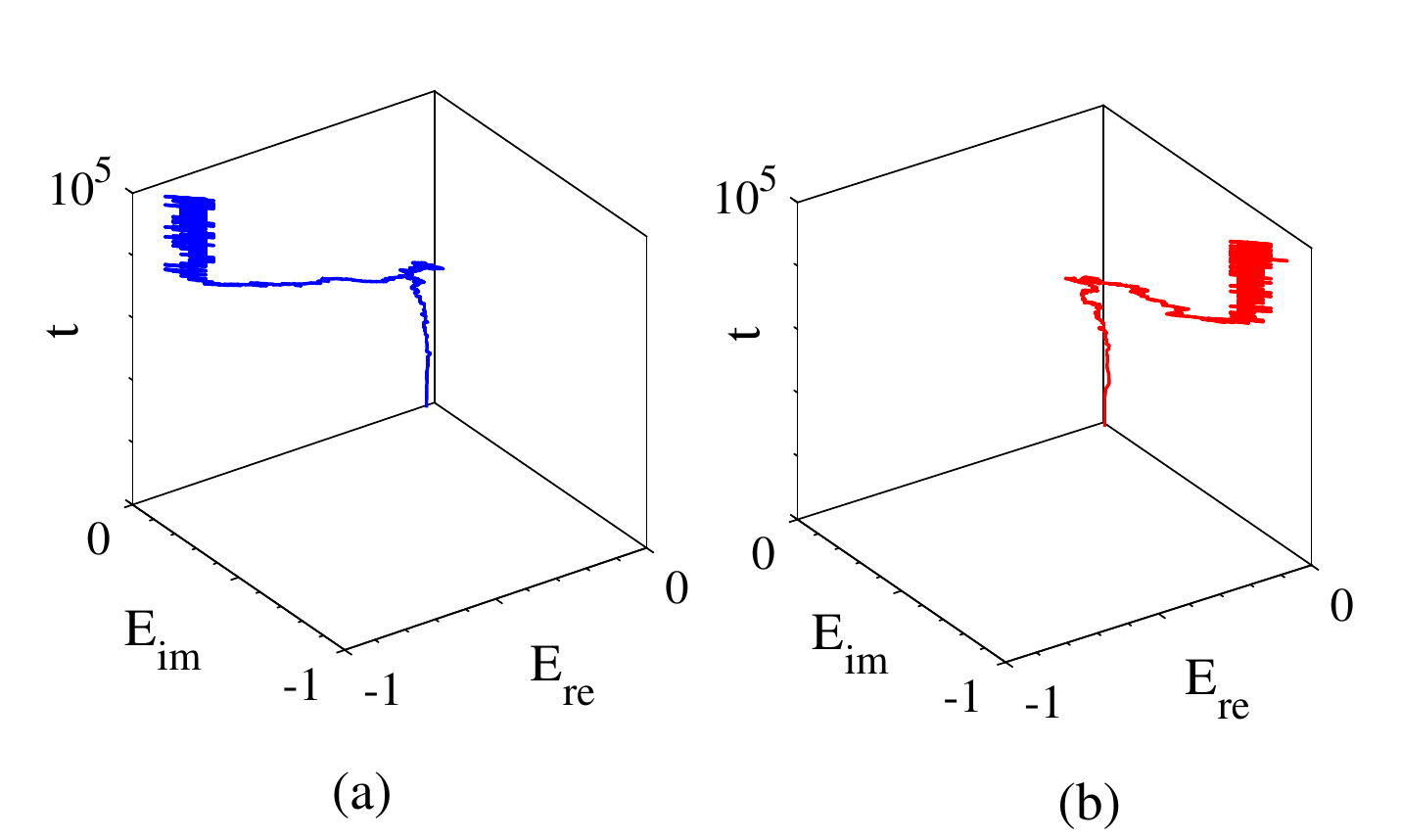}
	\caption{ three-dimensional plot, energy vs. time, before and after reaching a balanced state.
	 The share of $E_{re}$ and $E_{im}$ in total energy vs time for two arbitrary configurations which fall in the (a) real domain or (b) imaginary domain. After reaching a balanced state, the system switches between fixed points with the same magnitude of energy. The $t$ axis shows the Monte-Carlo steps as a time unit.} 
	\label{Fig5}
\end{figure}

Fig. \ref{Fig5} is a three-dimensional representation, which shows the share of $E_{re}$ and $E_{im}$ in total energy versus time for two arbitrary configurations. Unlike the Heider balance, which has only one balance point $(E=-1)$, competitive balance has multiple fixed points in the $E=E_{re}+E_{im}=-1$ line. As the figure shows, the energy path reaches a balanced state in one domain, then starts to switches between the fixed points with the same magnitude of energy in that domain. 

Indeed, the system evolves based on increasing the number of balanced triads, to the point that it reaches a balanced state. In this state, all the triads are balanced and the only flip that is acceptable is a flip that transforms one type of balanced triad to another.  Using this figure, we illustrate that the system does not stay in just one balanced fixed point, but switches between different available balanced values.

\subsection{Asymmetric cases}

\begin{figure}[!h]
      \includegraphics[width=1\linewidth]{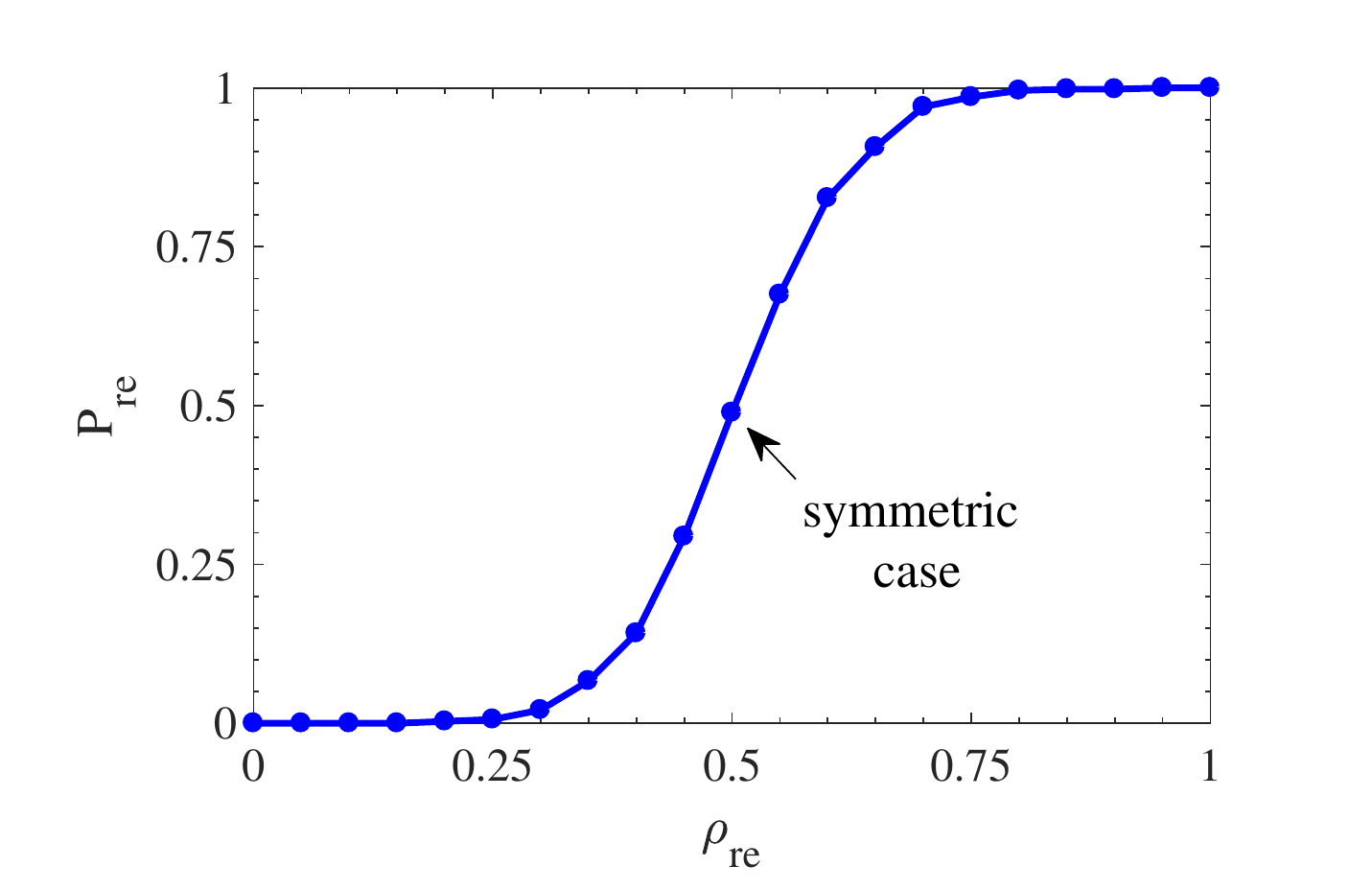}
	 \caption{The probability of falling in the real domain among all the realizations that reach to the global minima ($P_{re}$) as a function of the density of real edges in initial condition ($\rho_{re}$).}
	\label{Fig6}
\end{figure}

 In symmetric cases, both real and imaginary edges have equal chance to dominate the system over the evolution. If the system, however, becomes biased, then dominance of one of the interests become more probable. Asymmetric conditions can occur in various forms. Amongst them two forms could be more probable:

(1) Asymmetric initial condition: If the initial conditions are not symmetric i.e. the density of one type is higher than the other one, then the chance of dominance of the interests will not be equal. To estimate the role of asymmetric initial conditions we perform a simulation. An ensemble of networks with asymmetric initial conditions evolves over time. We then measure the probability of falling in the real domain  (among all the realizations that reach the global minima) as a function of the density of real edges in initial condition. The result of simulation has been graphed in Fig. \ref{Fig6}. 

(2) Another form of asymmetry can occur in weights of real and imaginary triads in the Hamiltonian: In this condition, real balanced and imaginary balanced triads have different shares in the Hamiltonian. In other words, the Hamiltonian can be defined as follows:
\begin{eqnarray}
\label{eq:8}
\centering
E=\alpha E_{re}+\beta E_{im},
\end{eqnarray}
where for asymmetric cases we suppose $\alpha\neq\beta$. While in symmetric cases dominances of both relations are equal, in asymmetric cases the network has only one global minimum where not only all triads are balanced but also all triads are of the real (imaginary) kind if $\alpha>\beta \; (\alpha<\beta)$, so there is no symmetry to be broken and the final state is predictable.

\subsection{Density of triads in final states}
A macro level analysis showed us that  the global behavior of the network leads to the formation of three possible domains, based on the final energy levels. We now turn to a micro level analysis, study the details of network structure, and detect the kinds of triads that will remain in each domain. Here, we leave aside the jammed states and follow the density of triads overtime for the paths which finally settle to the global minima of energy.

\begin{figure}[!h]
      \includegraphics[width=1\linewidth]{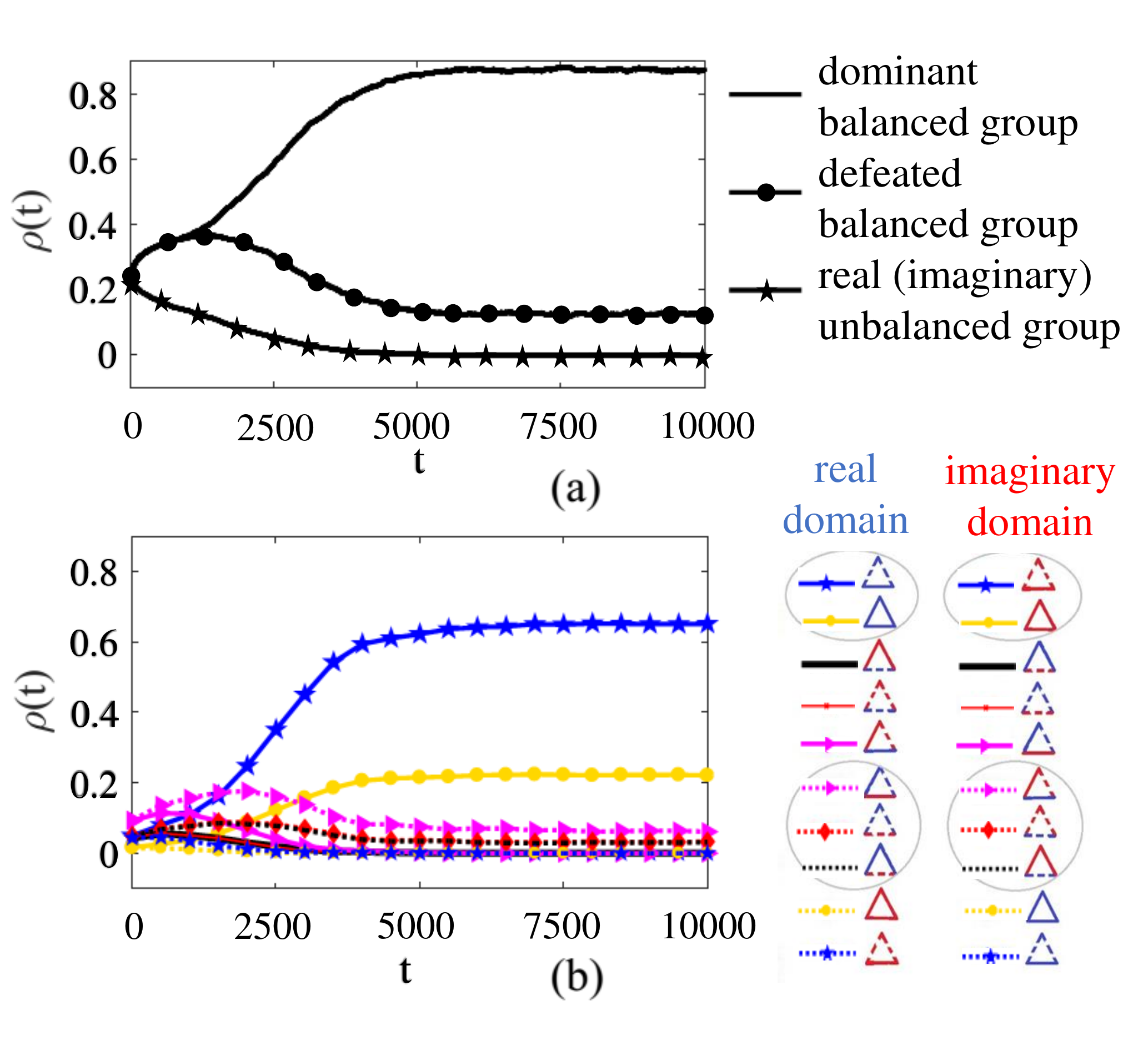}
	\caption{ The density of triads during the evolution of the network. 
		(a) The density of the dominant and defeated balanced and unbalanced groups across time. The unbalanced groups start to disappear from the beginning, and either the real or the imaginary balanced group dominates the network. (b) The density of different balanced triads across time. The legend shows the triads in the real or imaginary domains, in two separate columns. This means if the system leads to the real (imaginary) domain, the left (right) column is the related legend. The circled triads in the legend are the types that remain in the system in each specific domain.}
	\label{Fig7}
\end{figure}

As explained in Fig. \ref{Fig2}, the triads are classified into balanced and unbalanced, with real and imaginary energies. Since the initial densities of all types of edges are equal, as it is shown in Fig. \ref{Fig7}a, all four groups of triads have the same density at the beginning. Over the early steps of evolution, the number of balanced triads dominates unbalanced ones, resulting in division of branches in the figure. As time goes by, at a point, symmetry between the number of imaginary and real triads is spontaneously broken and one of the balanced groups predominates, while the density of the other balanced group decreases. After saturation the dominant balanced group occupies more than 80\% of all triads leaving less than 20\% for the defeated balanced group and nothing for unbalanced groups.

Another question is the share of different possible balanced triads in the final states. The density of balanced triads across time for realizations that lead to the real and imaginary domain is shown in Fig. \ref{Fig7}b, as indicated by the legends in the left and right column.  Surprisingly, we observed that  although constraints concerning energy allow the existence of ten forms of balanced triads only subgroups of them coexist in the final global states. So, if the system falls into the real domain, out of ten balanced triads only five of them remain: two pure real balanced triads and three mixed imaginary ones, which are specified by circles in the left column in Fig. \ref{Fig7}b. This form of selection indicates that it is more favorable to the system to go to the state with the most real links. In the same way, if the system gets into the imaginary domain [Fig. \ref{Fig7}b based on the right column legend], the densest triads are, respectively, pure imaginary and mixed real ones which have been identified in the right column. 
As a result, the steady state of a competitive system is where real or imaginary links overcome the other one. Indeed, in the real (imaginary) domain, real (imaginary) links predominate in the network and a few imaginary (real) links remain sporadic across them.

The point is that we have two groups of triads which belong to the real domain and the imaginary domain represented in Fig. \ref{Fig7}b. If we focus on triads in the real domain we observe that balanced triads of this domain can be switched by only one update of the connections. None of the balanced triads in the real domain can be turned to a balanced triad of the imaginary domains through a mere update. This means that to move from a balanced triad in the real domain to the imaginary domain energy needs to be added, which is not allowed in our dynamic. So, when the energy decreases in our dynamic once the number of triads of a domain meaningfully bypasses the number of triads in the other domain, then triads of this domain have the flexibility to switch to each other and help the system decrease its energy. Moreover, in the real domain, the number of real edges is bigger in balanced triads. Having these two properties helps the system choose one of the cases where the majority of edges are either real or imaginary.

\subsection{Size analysis}
\begin{figure}[!h]
       \includegraphics[width=1\linewidth]{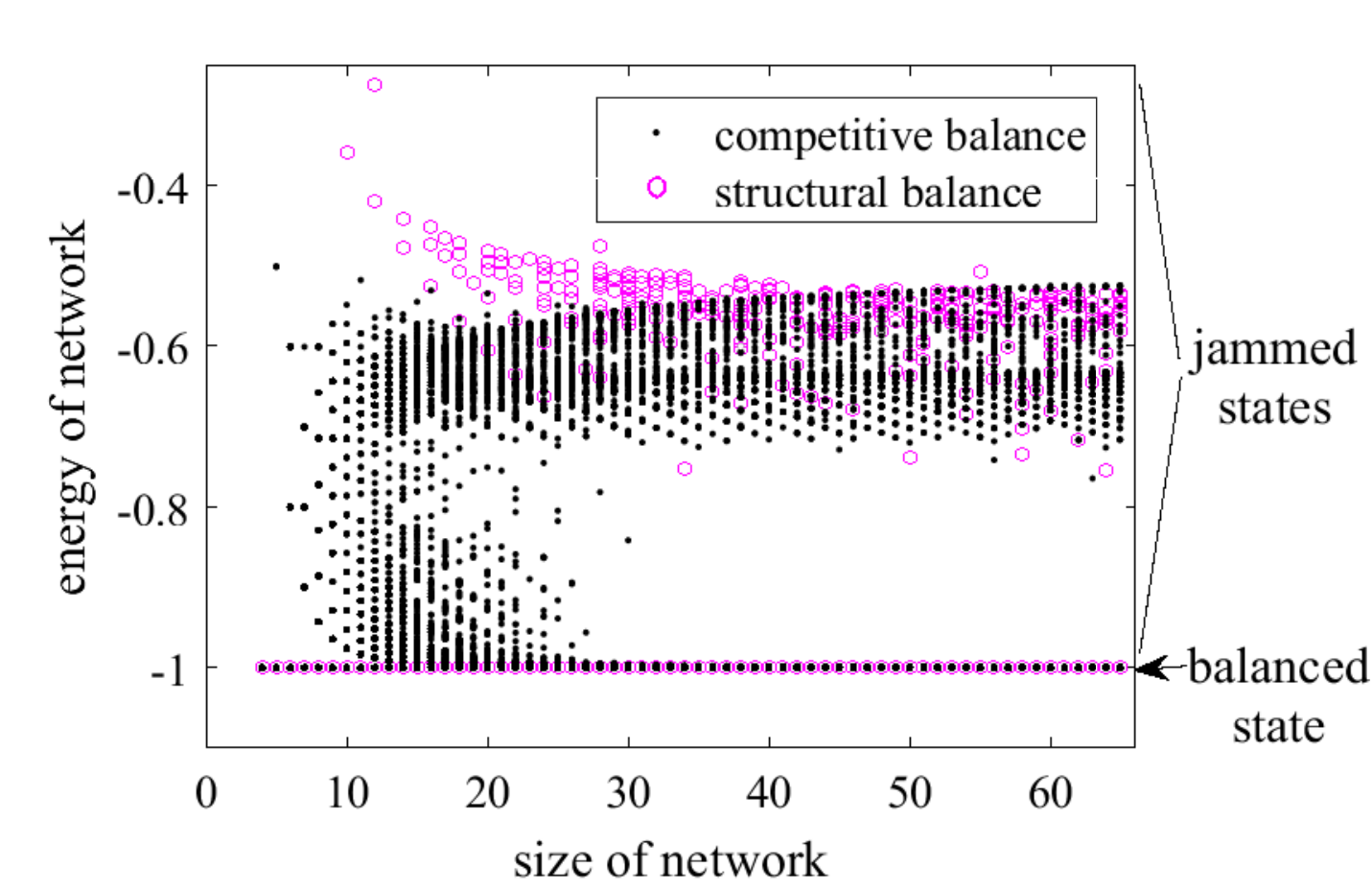}
	\caption{ Comparison between structural balance and competitive balance models, and  the final energy of 1000 realizations for different sizes of the
network. Each point represents the final energy of each realization.
Some points overlap in the figure, especially $E=-1$.}
	\label{Fig8}
\end{figure}

Now we investigate the effect of size on our results. We performed simulation for a wide range of sizes $3\leq N \leq64$ in ensembles of 1,000 iterations. We let the simulation run long enough. Before reaching a balanced state, if no update was observed to decrease the energy of the network, then a jammed state was counted. The final states which are shown in Fig. \ref{Fig8} are global and local minima of the energy.  In this figure, we compare the possible final states between structural balance and competitive balance models for different sizes of networks. Size analysis reveals that, although in Heider's balance model, networks of certain sizes (i.e. $N<9\; \&\;N=10$) have no jammed states \cite{Antal2005}, this is not the case in our model, and jammed states can occur in any size $N>3$ in the competitive balance model. 

\begin{figure}[!h]
      \includegraphics[width=1\linewidth]{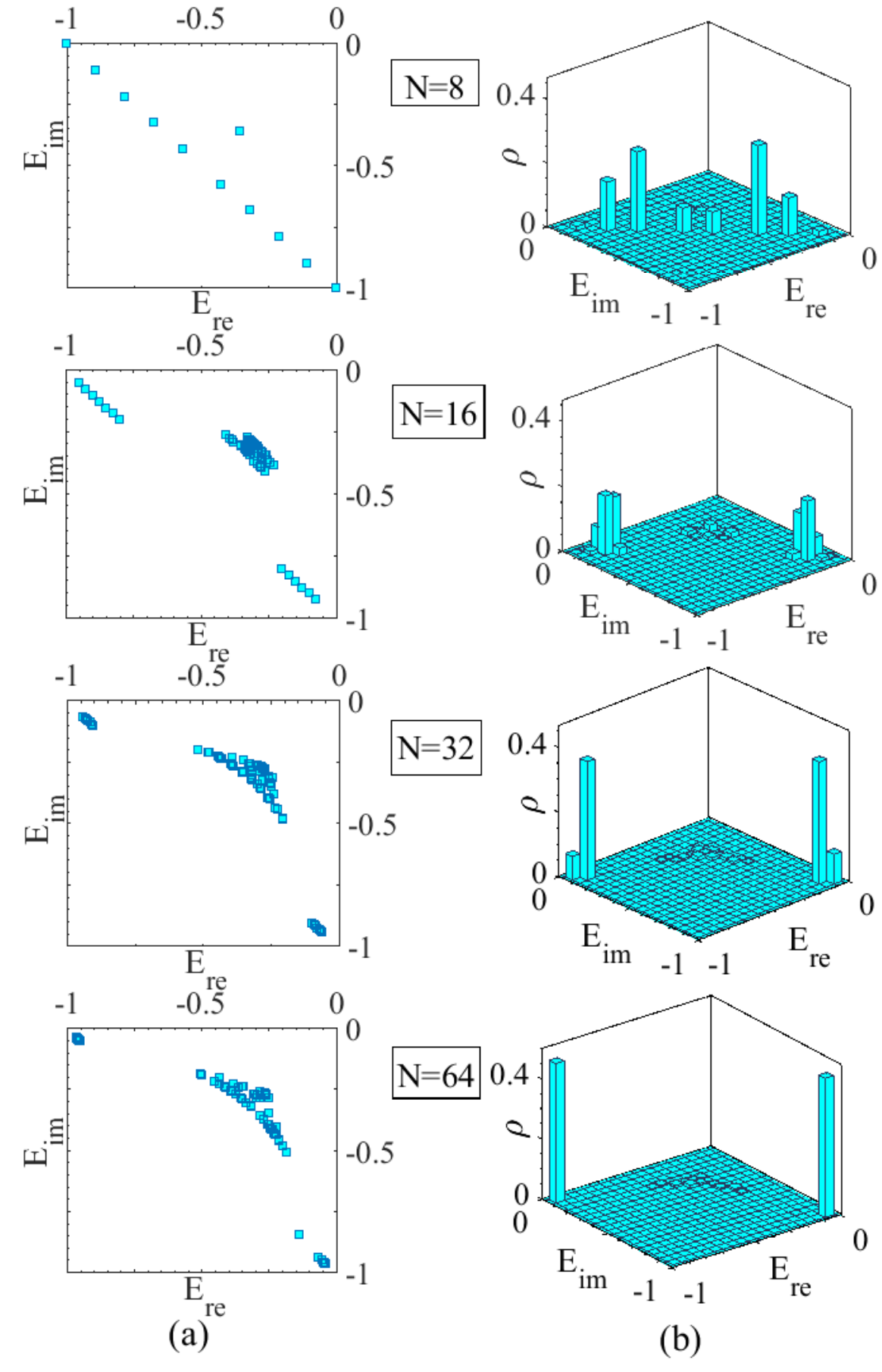}
	\caption{ Size effect in the final states.
		(a) The fixed points of different domains, and (b) the probability of occurrence of fixed points for networks with sizes $N=$ 8, 16, 32, and 64.}
	\label{Fig9}
\end{figure}

 Fig. \ref{Fig9}a illustrates the final energies for a selection of sizes, and Fig. \ref{Fig9}b shows the probability density for the spectra.
 Additionally, as the spectrum of final states in Fig. \ref{Fig9} shows, as the size grows the probable global minima move towards homogeneity. In other words final global minima get closer to the real or imaginary axes. This means that for the global minima falling in the real domain, the share of imaginary triads shrink as size grows. The same story happens if the system fall in the imaginary domain. In society this implies that in fully connected networks, for large sizes we approach a more homogeneous world where one of the discourses almost eliminates the other. 
Actually such a size effect can be explained from the entropy point of view.

Consider a bipolar network with "real" edges. Amongst 20 types of triads, three types (left column in Fig. \ref{Fig7}) are those types that appear in the real domain while having one "imaginary" edge. We call such an edge an anomalous edge. 

Suppose we start in a bipolar world where all edges are real. In such a case, in each triad one link can flip to an imaginary edge with an opposite sign while energy is not changed. Now, we ask which ratio of triads can have an anomalous edge. 
Since no two such anomalous edges can be incident on the same node (since they will then form an unbalanced triad), at most, a number of $N/2$ edges can flip. As each edge belongs to $(N-2)$ triads, in the most extreme case, $N(N-2)/2$ triads can have anomalous cases. This is why in a connected network, there are $N(N-1)(N-2)/6$ triads. So, the density of triads with an anomalous edge shrinks as the size grows and tends to zero for large networks. So the size effect in Fig. \ref{Fig9} can be explained. 

Now, another concern arises. What will be the density of the most probable triad for such a uniform world? In size $N=20$ we observed that out of ten possible triads only five survived after the evolution of the system. As well, we noticed that one form of triads outnumbered the others by the rate of 60\%, graphed in Fig. \ref{Fig7}b. One might wonder at what rate the most popular triad dominates the system in the thermodynamic limit.

\begin{figure}[!h]
      \includegraphics[width=1\linewidth]{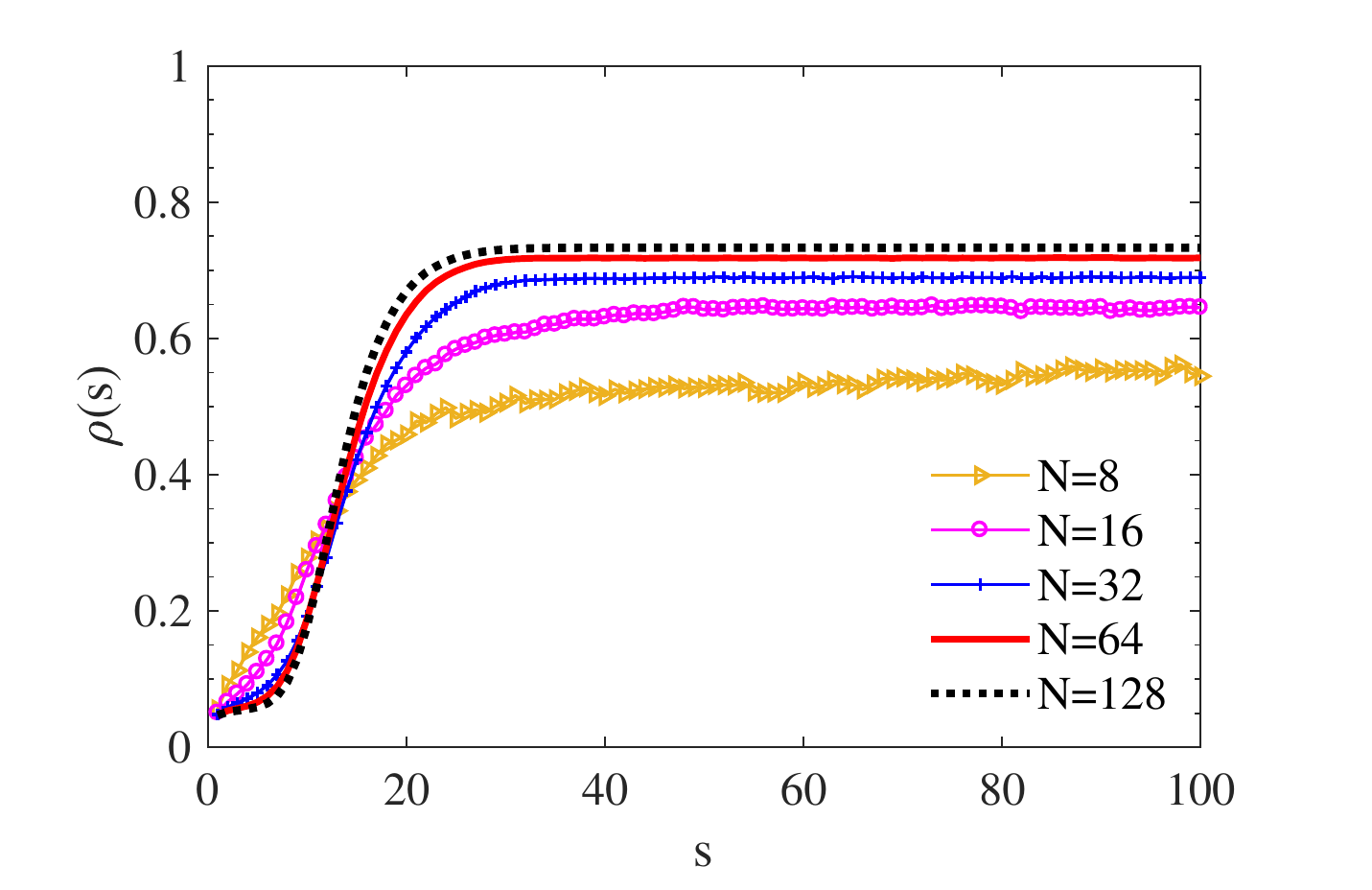}
	\caption{ Size effect in the triads density.
		Density of the most probable triad in the networks with sizes $N=$ 8, 16, 32, 64, and 128. The horizontal axis represents Monte-Carlo steps which are scaled by the number of edges for each network.}
	\label{Fig10}
\end{figure}

Fig. \ref{Fig10} graphs the evolution of the share of the most popular triads over time for various sizes. It seems that as the size grows the share of the most probable triad grows as well. However, an upper bound appears in the graph and is revealed by a rough calculation.  In a homogenized world, observed in larger sizes, with one of the discourses dominating the system, the configuration becomes either paradise or bipolar \cite{Cartwright1956}. Entropy, however, suggests the bipolar state, in which there are two strong subsets having intragroup friendly and intergroup hostile relationships. The number of possible choices in a bipolar network as such is maximized if each group has almost half the population \cite{Antal2005} $N/2$. In this system two kinds of triads are formed: about $N^3/24$ triads with edges $(1,1,1)$  and $N^3/8$ triads with edges $(1,1,-1)$. So, referring back to Fig. \ref{Fig10}, the $(1,1,-1)$ triad leads the density with an optimum share of $0.75$.

\section{\label{sec4} Conclusion }

Heterogeneity and conflict of interest are inherent in most real-life social networks. While a typical property of most social networks is their strive towards balance, as theorized by Heider's pioneering work, we discuss that different desired states of balance may coexist in a network, with one state working against others. In this paper, we introduce the competitive balance theory by redressing the original balance function and modifying the model to take the conflict of interest into account.  The competitive balance theory studies the evolution of  networks towards different states of balance, in the presence of heterogeneous and potentially competing types of  links. 

In this paper, we assumed that the system evolves towards two competing states of balance and comprises two different types of friendship and hostility. 
We used the complex plane to distinguish between the two different types without disturbing the homogeneity of triads in the Hamiltonian function. This way, while proposing a new model, we follow the original conceptualization of balanced and unbalanced triads by Heider. 

In structural balance dynamics, there is one single fixed point for the balanced final state; while our model follows the same dynamics, all the points on the $ E_{re}+E_{im}=-1$ line  are mathematically possible answers. However, a very limited number of these possible values are observed in our simulation. In fact, the introduction of different types of edges adds a constraint to the problem which makes many of the possible values unacceptable or highly unlikely. Our results show that in the final balanced state the system switches between different possible values in the reached domain. 

In our simulation, we randomly distribute four different edges, two real and two imaginary, to represent different types of friendship and hostility. As the network evolves, although at first there is no preference for the share of real or imaginary parts in energy, at a specific point there appears to be a symmetry breaking. The system, typically, moves towards the domination of the real or imaginary domain, while the nondominant type of energy maintains a marginal presence. It was interesting for us to see that whenever the system reaches a global minimum, one type of edge prevails. For a better comprehension of this state, imagine a normative discourse (e.g. heteronormative) staying\_or an alternative discourse (e.g. LGBTQ inclusive) becoming\_the dominant discourse, while the other maintains a marginal existence.

Size analysis shows that in smaller sizes the network demonstrates higher levels of tolerance for the nondominant type, though, as the network grows so does its intolerance for the marginal presence of the other discourse, while in the thermodynamic limit the nondominant type of links ceases to exist.
 
Size analysis shows that in smaller sizes the network demonstrates higher levels of tolerance for the nondominant type, though as the network grows so does its intolerance for the marginal presence of the other discourse, while in the thermodynamic limit the nondominant type of links ceases to exist.

Note that, in this paper, we take two different conflictual states of balance into account; this modeling choice, depending on the research question or the case under study, can be extended to more than two conflicting sets of ideals.



\end{document}